\documentclass[12pt,preprint]{aastex}
\usepackage{natbib}
\begin{document}
\slugcomment{ApJ in press 2013-10-25}
\hyphenation{mag-net-ic-al-ly}

\title{Herbig Stars' Near-Infrared Excess: An Origin in the\\
  Protostellar Disk's Magnetically-Supported Atmosphere}

\author{N.~J.\ Turner\altaffilmark{1,2},
  M.\ Benisty\altaffilmark{2,3},
  C.~P.\ Dullemond\altaffilmark{2,4} and
  S.\ Hirose\altaffilmark{5}
}
\altaffiltext{1}{Jet Propulsion Laboratory, California Institute of
  Technology, Pasadena, California 91109, USA;
  neal.turner@jpl.nasa.gov}
\altaffiltext{2}{Max-Planck-Institut f\"ur Astronomie, K\"onigstuhl
  17, D-69117 Heidelberg, Germany}
\altaffiltext{3}{Observatoire de Grenoble, Universit\'e Joseph Fourier,
  Grenoble, France}
\altaffiltext{4}{Institut f\"ur Theoretische Astrophysik, Zentrum
  f\"ur Astronomie, Universit\"at Heidelberg, Albert-Ueberle-Str.~2,
  69120 Heidelberg, Germany}
\altaffiltext{5}{Institute for Research on Earth Evolution, Japan
  Agency for Marine-Earth Science and Technology, 3173-25 Showamachi,
  Kanazawa-ku, Yokohama, Kanagawa 236-0001, Japan
}

\begin{abstract}
  Young stars with masses 2--8~Suns, called the Herbig~Ae and~Be
  stars, often show a near-infrared excess too large to explain with a
  hydrostatically-supported circumstellar disk of gas and dust.  At
  the same time the accretion flow carrying the circumstellar gas to
  the star is thought to be driven by magneto-rotational turbulence,
  which according to numerical MHD modeling yields an extended
  low-density atmosphere supported by the magnetic fields.  We
  demonstrate that the base of the atmosphere can be optically-thick
  to the starlight and that the parts lying near 1~AU are tall enough
  to double the fraction of the stellar luminosity reprocessed into
  the near-infrared.  We generate synthetic spectral energy
  distributions (SEDs) using Monte Carlo radiative transfer
  calculations with opacities for sub-micron silicate and carbonaceous
  grains.  The synthetic SEDs closely follow the median Herbig SED
  constructed recently by Mulders and Dominik, and in particular match
  the large near-infrared flux, provided the grains have a mass
  fraction close to interstellar near the disk's inner rim.
\end{abstract}

\keywords{protoplanetary disks --- radiative transfer}

\section{INTRODUCTION\label{sec:introduction}}

Young intermediate-mass stars with disks commonly show a large
near-infrared excess at wavelengths 2--4~$\mu$m.  The luminosity in
the excess is often a noticeable fraction of the stellar bolometric
luminosity.  Hydrostatic disk models fail to reproduce the
near-infrared flux, emitting too little by a factor two for some stars
\citep{2006ApJ...636..348V}.  This is a puzzle because in the basic
hydrostatic picture each star is allowed only a narrow range of
fluxes, determined as follows \citep{2010ARA&A..48..205D}: the
2-$\mu$m emission arises in material with temperatures near the
silicate sublimation threshold of about $1\,500$~K.  Such temperatures
occur at a certain distance from the star, inside which the dust
cannot survive, leaving the disk optically-thin.  The material at the
sublimation front therefore faces the star directly, intercepting and
re-emitting at 2~$\mu$m a fraction of the stellar luminosity equal to
the ratio of the disk thickness to the front's radius.  The disk's
density scale height in hydrostatic balance is proportional to the
sound speed divided by the orbital frequency, two quantities that both
are fixed --- the sound speed by the sublimation temperature, and the
orbital frequency by the radius where the sublimation temperature is
reached.  Varying the grain size and composition shifts the
sublimation front somewhat, but the ratio of density scale height to
sublimation radius changes little.  For purposes of the near-infrared
excess, the disk's thickness is equal to the height of the surface
where the starlight is absorbed and reprocessed.  This
starlight-absorbing surface lies a few density scale heights from the
equatorial plane, growing logarithmically with the surface density
owing to the steep density profile of the hydrostatic disk atmosphere.
In quite a few objects, the near-infrared excess cannot be accounted
for even with extreme disk masses.

The puzzle only grows stronger when longer infrared wavelengths are
considered.  Matching the excess at 7~$\mu$m and the 13.5-to-7-$\mu$m
flux ratio requires artificially scaling up the disk's thickness near
1~AU by factors of as much as three compared with hydrostatic models,
in the sample of 33~Herbig stars examined by
\citet{2009A&A...502L..17A}.

Two main ideas have been proposed to explain the large near-infrared
excesses.  The first is that the disk is thicker because it is hot.
The extra heating might come from accretion.  Herbig systems'
H$\alpha$ equivalent widths, a measure of the accretion luminosity,
correlate with the ratio of the K-band excess flux to that in the
H-band, suggesting a link between accretion and the inner disk shape
\citep{2006ApJ...653..657M}.  However as Manoj et al.\ were aware, the
accretion power by itself cannot explain the observed near-infrared
fluxes without mass flow rates exceeding $10^{-6}$~$M_\odot$~yr$^{-1}$
\citep{1992ApJ...397..613H, 1992ApJ...393..278L}.  Yet the flow rate
must be less than $10^{-7}$~$M_\odot$~yr$^{-1}$ for the gas inside the
silicate sublimation radius to be optically thin
\citep{1993ApJ...407..219H} and compatible with the central holes
detected in near-infrared interferometric observations of many Herbig
stars.  The holes have sizes generally consistent with the sublimation
radius over a broad range of stellar luminosities, as reviewed by
\citet{2010ARA&A..48..205D}.  Alternatively, the extra heating might
come from fast-moving electrons ejected from the grains by stellar
ultraviolet photons.  At the low gas densities found in the
hydrostatic atmosphere, heat transfer from gas to dust is inefficient.
The photoelectron-heated gas cannot easily cool, and reaches
temperatures up to several thousand degrees
\citep{2011MNRAS.412..711T}.  However the low gas densities also mean
that even sub-micron grains quickly settle out, so the hot material is
likely to be transparent to the starlight.

The other main idea is an extra system component that is warm enough
to emit significantly in the near-infrared, such as a spherical halo
or envelope \citep{1993ApJ...407..219H, 2006ApJ...636..348V}, dusty
disk wind \citep{2007ApJ...658..462V, 2012ApJ...758..100B} or cloud of
dust ejected by magnetic forces \citep{2012ApJ...745...60K}.  A
component covering a large solid angle might help account for the
common occurrence of variable circumstellar extinction among the
Herbig~Ae stars \citep{1991A&AS...89..319B, 1998A&A...331..211M,
  2000prpl.conf..559N} and especially the members of the UX~Ori class
\citep{1998AstL...24..802G, 2009AstL...35..114G}.  However spherical
structures by themselves appear incompatible with interferometric
measurements \citep{2001Natur.409.1012T, 2004ApJ...613.1049E}.
Combining a spherical halo of modest optical depth with a hydrostatic
disk yields a better fit than either component alone in several cases
\citep{2006ApJ...647..444M, 2011A&A...528A..91V, 2012A&A...541A.104C}.

Here we focus on the expectation that the accretion stresses in the
sufficiently-ionized parts of the disks come from magnetic forces
\citep{1988PThPS..96..151U, 1991ApJ...376..214B, 1996ApJ...457..355G,
  2009ApJ...701..620Z, 2011ARA&A..49..195A}.  A magnetically-supported
atmosphere is a natural consequence, since shearing-box MHD
calculations extending more than a few scale heights from the midplane
show magnetic fields generated in magneto-rotational turbulence and
rising buoyantly to form an atmosphere in which magnetic pressure
exceeds gas pressure \citep{2000ApJ...534..398M, 2010MNRAS.409.1297F}.
The atmosphere is optically-thin to its own continuum emission when
its base is set by the penetration of the young star's ionizing X-ray
photons \citep{2009ApJ...701..737B}, but is nevertheless
optically-thick to the starlight which illuminates the disks around
low-mass T~Tauri stars at grazing incidence
\citep{2011ApJ...732L..30H}.

In this contribution we demonstrate that the magnetically-supported
atmosphere is optically-thick to the starlight in Herbig disks too.
The magnetic support can make the inner disk two to three times
taller, so that it intercepts and reprocesses into the near-infrared a
correspondingly greater fraction of the stellar luminosity.

Our model star is described in section~\ref{sec:star} and the disk
supported jointly by gas and magnetic pressure in
section~\ref{sec:disk}.  After choosing dust opacity curves
(section~\ref{sec:dust}) we use a Monte Carlo radiative transfer
approach (section~\ref{sec:transfer}) to compute the disk's shape and
temperature, solving jointly for global radiative and vertical
magneto-hydrostatic equilibrium (section~\ref{sec:hse}).  Synthetic
observations are made using the ray-tracing method described in
section~\ref{sec:raytracing}.  The results are set out in
section~\ref{sec:results} and the summary and conclusions follow in
section~\ref{sec:conclusions}.

\section{STAR\label{sec:star}}

Our star is a 2.4-$M_\odot$ Herbig~Ae modeled on AB~Aurigae.  Its
radius $2.55 R_\odot$ and temperature $9\,550$~K yield luminosity
47.9~$L_\odot$ \citep{1998A&A...330..145V}.  The star emits the
spectrum of the Solar-metallity Kurucz model with the nearest gravity
and effective temperature --- $10^4$~cm~s$^{-2}$ and $9\,500$~K.

\section{DISK\label{sec:disk}}

\begin{figure}[tb!]
  \epsscale{0.7} \plotone{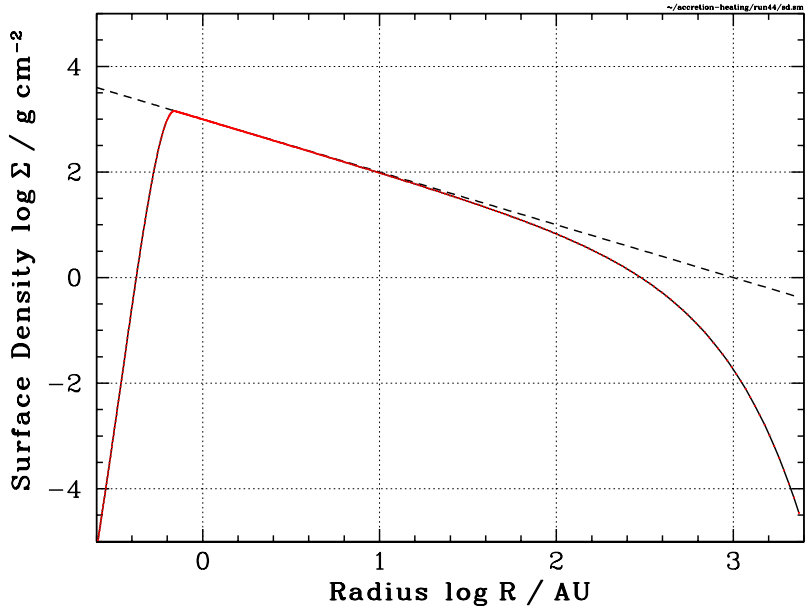} \figcaption{\sf Surface density
    profile in the model Herbig disks (solid curve).  The radiative
    transfer grid cells' radial edges are marked by tiny red squares.
    The dashed black diagonal line shows a $1/r$ dependence.
    \label{fig:sd}}
\end{figure}

We assume the surface density $\Sigma$ falls inversely with radius $r$
until cut off exponentially at outer radius $r_o$:
\begin{equation}\label{eq:sd}
  \Sigma = {M_d\over 2\pi r_o^2} \left(r_o\over r\right)
  \exp\left(-r/r_o\right).
\end{equation}
This is a similarity solution with total mass $M_d$, obtained under a
simple viscosity prescription \citep{1998ApJ...495..385H}.
Eq.~\ref{eq:sd} also is a fair match to the surface densities measured
at separations of tens of~AU using millimeter interferometry of the
dust continuum emission from T~Tauri stars
\citep{2010ApJ...723.1241A}.  We set the outer cutoff radius to
$r_o=250$~AU.  In addition we cut the disk short inside a radius
$r_i$.  This is meant to model not the stellar magnetosphere's
truncation of the gas, but the sublimation front's truncation of the
optical depth.  Directly solving for the position of the front can
introduce convergence issues \citep{2009A&A...506.1199K} which we wish
to avoid.  Inside $r_i$ we roll off the surface density by the factor
$\exp\left[-\left(\left\{r-r_i\right\}/\Delta r_i\right)^2\right]$.
The inner cutoff radius and scale length are $r_i=0.7$ and $\Delta
r_i=0.1$~AU.  With these choices the disk is optically-thin inside
about 0.4~AU which is near the expected sublimation radius for our
model star.  We check after the fact that the temperature at unit
radial optical depth is close to the sublimation threshold.  Solving
in detail for the shape of the sublimation front is unlikely to
significantly reduce the fraction of the starlight intercepted by the
disk within 1~AU, but could change how the reprocessed luminosity is
distributed across near-infrared wavelengths.  The surface density is
$1\,000$~g~cm$^{-2}$ at 1~AU, yielding a total disk mass $0.176
M_\odot$ or 7\% of the stellar mass.  The surface density profile,
shown in figure~\ref{fig:sd}, along with the temperatures found as
described below, yields a Toomre $Q$ parameter that is smallest near
160~AU, where it exceeds seven in all models, indicating stability
against self-gravity.

\begin{figure}[tb!]
  \epsscale{0.7} \plotone{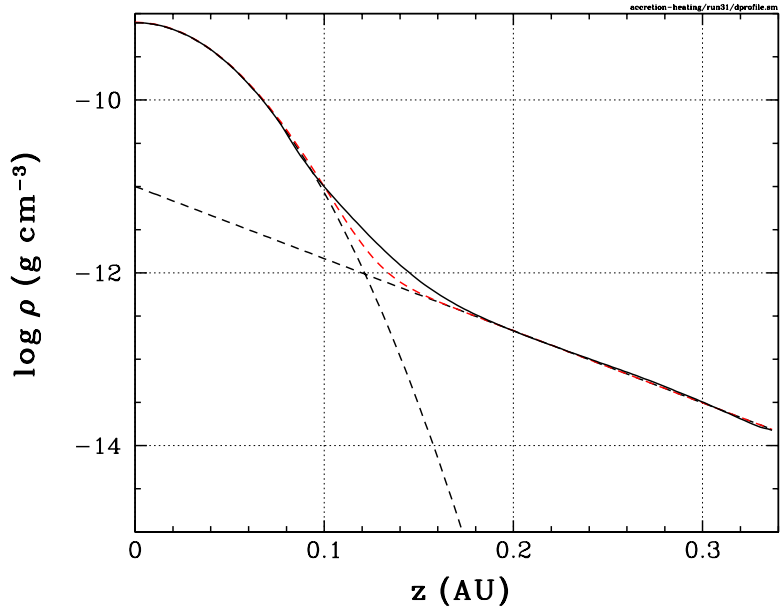} \figcaption{\sf Mean density profile
    in the shearing-box radiation-MHD calculation of
    \cite{2011ApJ...732L..30H} (solid black line).  The profile is fit
    by summing two components, a Gaussian and an exponential
    (eq.~\ref{eq:dprofile}; dashed black lines) with the total shown
    by a dashed red curve.  The fit is good except near the transition
    between the components, where the MHD result is up to 60\% denser.
    \label{fig:dprofile}}
\end{figure}

The vertical profile of density in each disk annulus is obtained as
follows.  First, we fit the mean density profile in the fiducial
radiation-MHD calculation from \cite{2011ApJ...732L..30H} with the sum
of a gas-pressure-supported isothermal interior and a
magnetically-supported atmosphere (figure~\ref{fig:dprofile}).  The
interior has a Gaussian density profile in the height $z$, while the
atmosphere is exponential:
\begin{equation}\label{eq:dprofile}
  {\rho(z)\over \rho_0} = \exp\left(-{z^2\over 2H^2}\right)
  + \frac{1}{78.6} \exp\left(-{z\over 1.57 H}\right),
\end{equation}
where $\rho_0$ is the Gaussian's midplane density, $H = c_s(z=0) /
\Omega$ the density scale height, $\Omega$ the Keplerian orbital
frequency and $c_s(z=0)$ the midplane isothermal sound speed, computed
using the gas mean molecular weight~2.3.

In the Monte Carlo radiative transfer calculations we use $H$ to
rescale the MHD results which \citet{2011ApJ...732L..30H} obtained
1~AU from an 0.5-M$_\odot$ T~Tauri star.  To be conservative we round
down the exponential scale length and normalization, making the
magnetically-supported atmosphere a little more compact and
lower-mass.  We adopt the density profile
\begin{equation}\label{eq:dprofile2}
  \rho(z) = \rho_{HSE}(z) + N\rho_0\exp\left(-{z\over AH}\right),
\end{equation}
where $\rho_{HSE}$ is the profile obtained by solving for vertical
hydrostatic balance while fixing the variation of the temperature with
mass column to the profile found in the previous Monte Carlo transfer
iteration.  The parameters are $N=1/80$, which normalizes the
exponential relative to the Gaussian, and $A=1.5$, the exponential
scale height in units of~$H$.  The hydrostatic component's midplane
density $\rho_0=\rho_{HSE}(0)$ is chosen using simple root-finding by
bisection, so that the total profile has the desired surface density.

We now consider three configurations: (A) the entire disk is supported
in the traditional fashion by gas pressure alone, (B) gas and magnetic
pressures contribute throughout according to eq.~\ref{eq:dprofile2},
and (C) magnetic pressure adds to gas pressure only in annuli with a
narrow range of radii just outside the silicate sublimation front,
where the high temperatures ensure good magnetic coupling through
collisional ionization of the alkali metals.  The three configurations
are listed in table~\ref{tab:support}.  The magnetically-supported
``bump'' in the third configuration is merged smoothly with the
surrounding gas-pressure-supported disk by giving the exponential
atmosphere's scale height a Gaussian radial variation about the
maximum value $A(r_b)=1.5$.  The Gaussian's FWHM is equal to the
radius $r_b$ of the bump's peak.

The atmosphere's thickness likely also depends on the net vertical
magnetic flux, which is a product of the global transport of magnetic
fields.  Given the uncertainties regarding this transport, we here
simply fix the atmosphere's scale height in units of the gas pressure
scale height to a value similar to that found by
\citet{2011ApJ...732L..30H}.  Their radiation-MHD calculations have a
net vertical magnetic flux with pressure $3\times 10^5$ times less
than the midplane gas pressure.

\begin{deluxetable}{lll}
  \tablecaption{\sf The three disk configurations' support against
    vertical gravity.  \label{tab:support}}
  \tablehead{\colhead{Name} & \colhead{Gas Support} & \colhead{Magnetic Support}}
\startdata
Gas           & Yes     & None \\
Magnetic      & Yes     & Throughout \\
Magnetic bump & Yes     & Only in bump near inner rim\tablenotemark{a} \\
\enddata
\tablenotetext{a}{The bump is centered $r_b=1$~AU from the star.}
\end{deluxetable}

\section{DUST OPACITY\label{sec:dust}}

\begin{figure}[tb!]
  \epsscale{0.7} \plotone{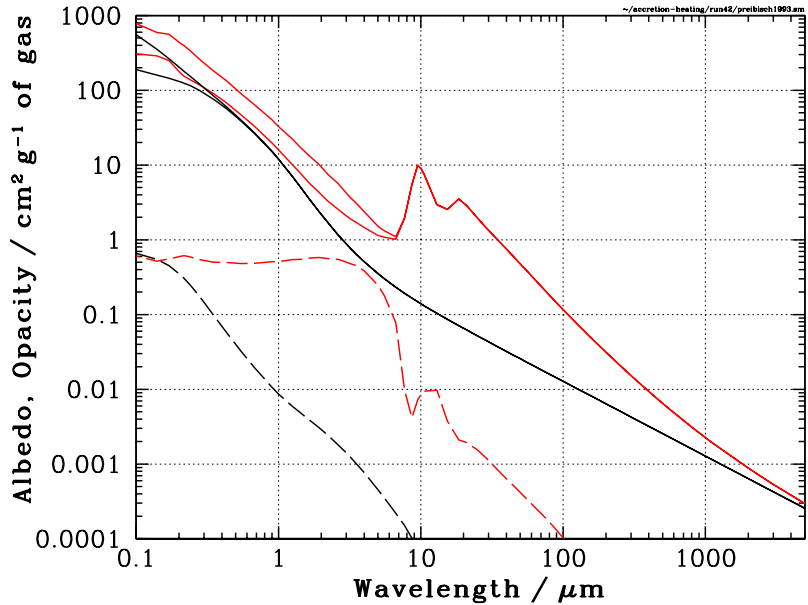} \figcaption{\sf Wavelength
    dependence of the opacities for the silicate and carbon particles
    (red) and for the carbon particles alone (black) from
    \citet{1993A&A...279..577P}.  The lower in each pair of solid
    curves is the absorption opacity, the upper the sum of absorption
    and scattering opacities.  Dashed curves indicate the
    corresponding albedos.  The two black solid curves nearly coincide
    due to the carbon particles' low albedos.
    \label{fig:opacities}}
\end{figure}

The disks' opacity comes primarily from dust, which we assume is
well-mixed in the gas except where otherwise specified.  We adopt
opacities from \cite{1993A&A...279..577P} who matched Mie calculations
of dust particles' optical response against data from molecular
clouds.  The grain model consists of silicate and carbonaceous
particles, each with a power-law size distribution of exponent $-3.5$.
The minimum and maximum sizes are 0.04 and 1~$\mu$m for the silicate
particles and 0.007 and 0.03~$\mu$m for the carbon particles.  The
opacity curves are shown in figure~\ref{fig:opacities} together with
the albedos, or ratios of scattering to total opacity.  Scattering
contributes about half of the total cross-section at optical
wavelengths, and is assumed isotropic.

SED modeling suggests dust is depleted in T~Tauri disk atmospheres by
factors 10--$10^4$ compared with the interstellar medium
\citep{2006ApJS..165..568F, 2009ApJ...703.1964F}.  A gas-to-dust ratio
of $12\,800$, in the same range, appears to be needed to understand
the water emission from T~Tauri stars \citep{2009ApJ...704.1471M}.
Furthermore, planet formation requires incorporating some of the solid
material into bigger bodies.  We therefore consider two dust-to-gas
mass ratios: the nominal interstellar value, as shown in
figure~\ref{fig:opacities}, and a hundredfold depletion.  The dust
opacities are simply scaled down by the depletion factor $\epsilon$.
Additionally we consider a scenario in which the dust takes the
depleted abundance except in a ring around $r_r=1\ {\rm AU}$, where
the peak dust mass fraction matches that in the dusty scenario.  The
mass fraction is a Gaussian in radius, asymptoting to the depleted
value far from the star.  The three dust distributions are listed in
table~\ref{tab:dust}.

\begin{deluxetable}{ll}
  \tablecaption{\sf The three dust distributions.  \label{tab:dust}}
  \tablehead{\colhead{Name} & \colhead{Dust-to-Gas Mass Ratio, $0.01\epsilon$}}
\startdata
Depleted   & $10^{-4}$ throughout \\
Dusty      & $10^{-2}$ throughout \\
Dusty ring\tablenotemark{a} &
$10^{-4}\times\left(1+99\exp\left[-4\left\{r/r_r-1\right\}^2\right]\right)$\\
\enddata
\tablenotetext{a}{The dusty ring is centered $r_r=1$~AU from the star.}
\end{deluxetable}

To summarize, each model disk is uniquely specified by listing the
magnetic support from table~\ref{tab:support} and the dust
distribution from table~\ref{tab:dust}.  Since we consider three
magnetic configurations and three dust scenarios, there are nine
models in all.

\section{RADIATION FIELD AND TEMPERATURE\label{sec:transfer}}

We compute the radiative equilibrium temperatures by emitting a large
number of photon packets from the star into the disk where they are
scattered, absorbed and re-emitted as many times as needed till they
escape to infinity.  With this approach the energy is conserved
exactly.  The stellar luminosity is divided equally among the packets.
We use the temperature relaxation procedure of
\cite{2001ApJ...554..615B}, drawing the frequencies of the re-emitted
packets from the difference between the old and new emission spectra,
such that the local radiation field adjusts to the updated
temperature.  The gas and dust are assumed to share a single
temperature at each point.  For efficiency when estimating the
radiation absorption rates and the radiation's mean intensity, we
include the contributions from all along the packet paths
\citep{1999A&A...344..282L}.  The intensity is accumulated in
20~contiguous non-overlapping wavelength bins including those centered
on the photometric bands U, B, V, R, I, J, H and K, the four Spitzer
IRAC channels at 3.6, 4.5, 5.8 and 8~$\mu$m, and the Spitzer MIPS
24-$\mu$m channel.  For contiguous wavelength coverage these are
rounded out with bands centered at wavelengths 1, 1.9, 2.79, 10, 12,
14.5 and 18~$\mu$m.  The center of each intensity bin appears as an
open circle on the spectral energy distributions in
section~\ref{sec:seds} below.

We compute the temperatures neglecting accretion heating.  Under
magneto-rotational turbulence, much of the released gravitational
energy is deposited in the disk atmosphere, at low optical depths to
the disk's own radiation \citep{2009ApJ...701..737B,
  2011ApJ...732L..30H}.  Heating at low optical depths has a reduced
effect on the midplane temperatures.  Also, the accretion heating
falls off with radius faster than the stellar irradiation heating, so
that including the accretion power would increase the disk's thickness
most near the inner rim.  By neglecting the accretion heating we thus
obtain a lower limit on the fraction of the stellar luminosity
reprocessed into the infrared near the disk's inner rim.

Because we neglect the accretion heating, the disk's interior is
isothermal on cylinders.  We therefore save the expense of computing
temperatures in the most optically-thick regions by simply bouncing
back any packets reaching a certain mass column, chosen so the
overlying material is optically thick at wavelengths near its thermal
emission peak.  The bouncing threshold is set to
$30/\epsilon$~g~cm$^{-2}$ in all the calculations shown here.  We
replace the missing interior temperatures by the mean of the last few
well-sampled values above.

The disk is divided into a grid of $N=800$~cells spanning the four
decades in radius from $r_0=0.25$ to $r_N=2500$~AU.  The cells are
concentrated near the inner rim to better resolve the transition from
optically-thin to thick in the radial direction.  Cell $j$'s inner
radius $r_j$ is given by $\log(r_j/r_0) = C\left[(\Delta/C+1)^{j/N} -
  1\right]$, where $\Delta=4$ is the log of the ratio of the grid's
outer to inner radius.  We choose a concentration parameter $C=0.3$.
The strength of the concentration can be seen in figure~\ref{fig:sd},
where dots mark the cell edges $r_j$.  In the vertical direction the
grid has 280~cells uniformly-spaced between the equatorial plane and
height $z=0.7 r_j$, yielding a spacing 0.25\% of the radius.  This
choice of upper boundary ensures all the material with starlight
optical depth greater than $10^{-3}$ lies on the grid, even in the
flared outer parts of our nine model disks.

\section{JOINT RADIATIVE AND MAGNETO-HYDROSTATIC EQUILIBRIUM
  \label{sec:hse}}

Once new temperatures have been found through the Monte Carlo
radiative transfer procedure, we restore vertical equilibrium by
reconstructing the density profile within each disk annulus as in
section~\ref{sec:disk}, while holding fixed the surface density and
the variation of the temperature with the mass column.

We iterate five times between radiative transfer and
magneto-hydrostatic balancing.  In each case the third, fourth and
fifth iterations show only minor differences, indicating the solution
is close to converged.  Each iteration of the radiative transfer
calculation involves $10^7$~photon packets.

\section{SYNTHETIC IMAGES AND SPECTRA\label{sec:raytracing}}

The procedure outlined above yields the density $\rho(r,z)$,
temperature $T(r,z)$ and frequency-dependent mean radiation intensity
$J_\nu(r,z)$.  Note that $J_\nu$ is taken piecewise constant across
each of the 20~wavebands described in section~\ref{sec:transfer}, the
individual photon packets' wavelengths having been discarded during
the accumulation.  From these three quantities we compute synthetic
images and spectra by solving the transfer equation,
\begin{equation}
\label{eq:transfer}
{dI_\nu\over d\tau_\nu} = I_\nu - {\kappa_\nu B_\nu(T) + \sigma_\nu J_\nu
\over \kappa_\nu + \sigma_\nu},
\end{equation}
on a grid of parallel rays extending toward the observer at infinity,
following \cite{1986arh..conf..141Y}.  The symbols have their usual
meanings \citep{1978stat.book.....M} with $\nu$ the frequency, $I_\nu$
the specific intensity, $J_\nu$ its angle average, $\tau_\nu$ the
optical depth and $\kappa_\nu$ and $\sigma_\nu$ the absorption and
scattering opacities.  The differential optical depth $d\tau_\nu$ over
a step of length $dl$ is $(\kappa_\nu+\sigma_\nu)\rho\,dl$.

Solving the transfer equation in this way is preferable over binning
the Monte Carlo photon packets in angle as they emerge from the
system, because it yields images with adequate spatial resolution
using far fewer packets.

\section{RESULTS\label{sec:results}}

\subsection{Does the Atmosphere Absorb Starlight?}

\begin{figure}[tb!]
   \centering
   \includegraphics[width=0.58\linewidth]{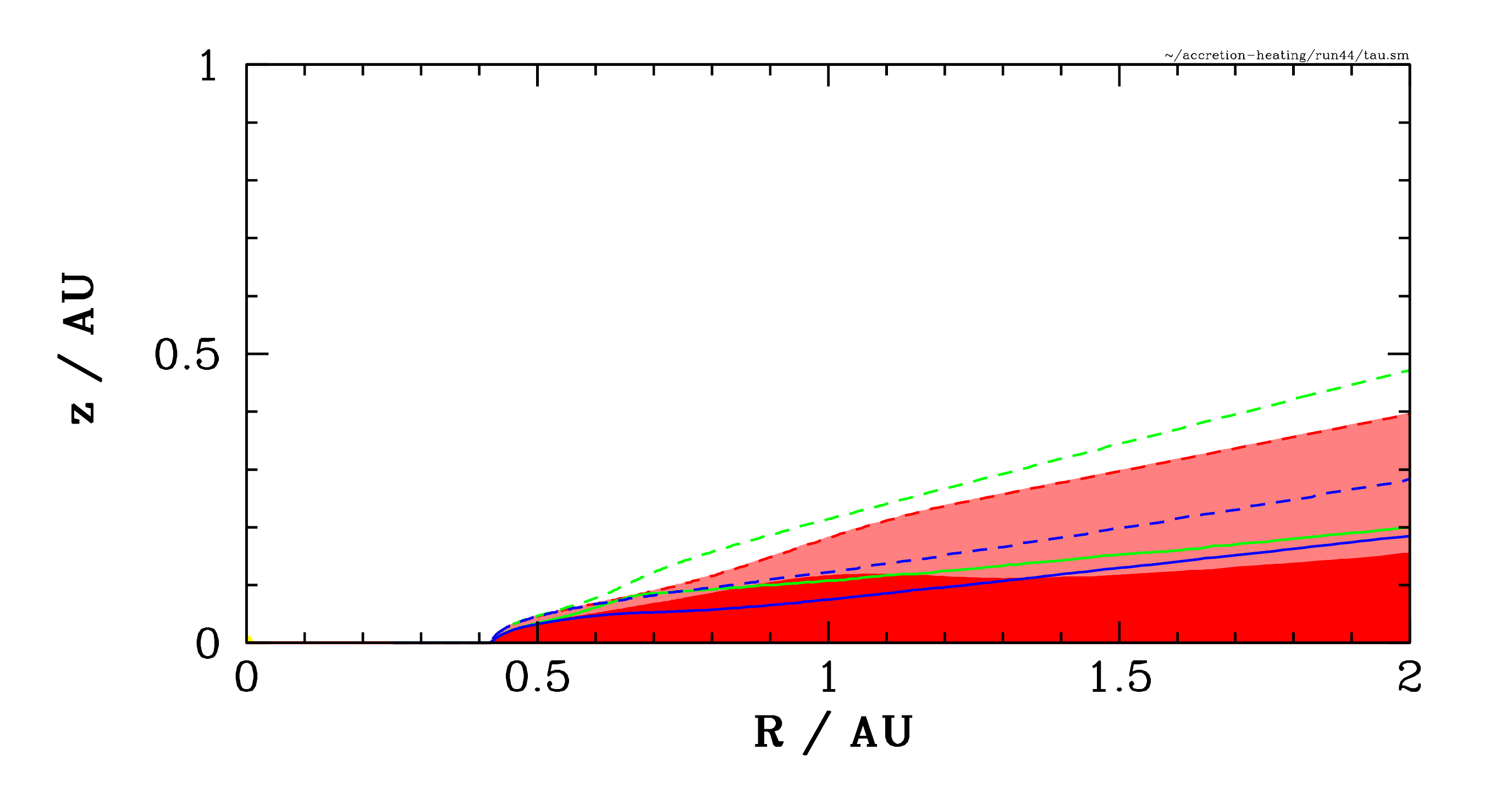}\\
   \includegraphics[width=0.58\linewidth]{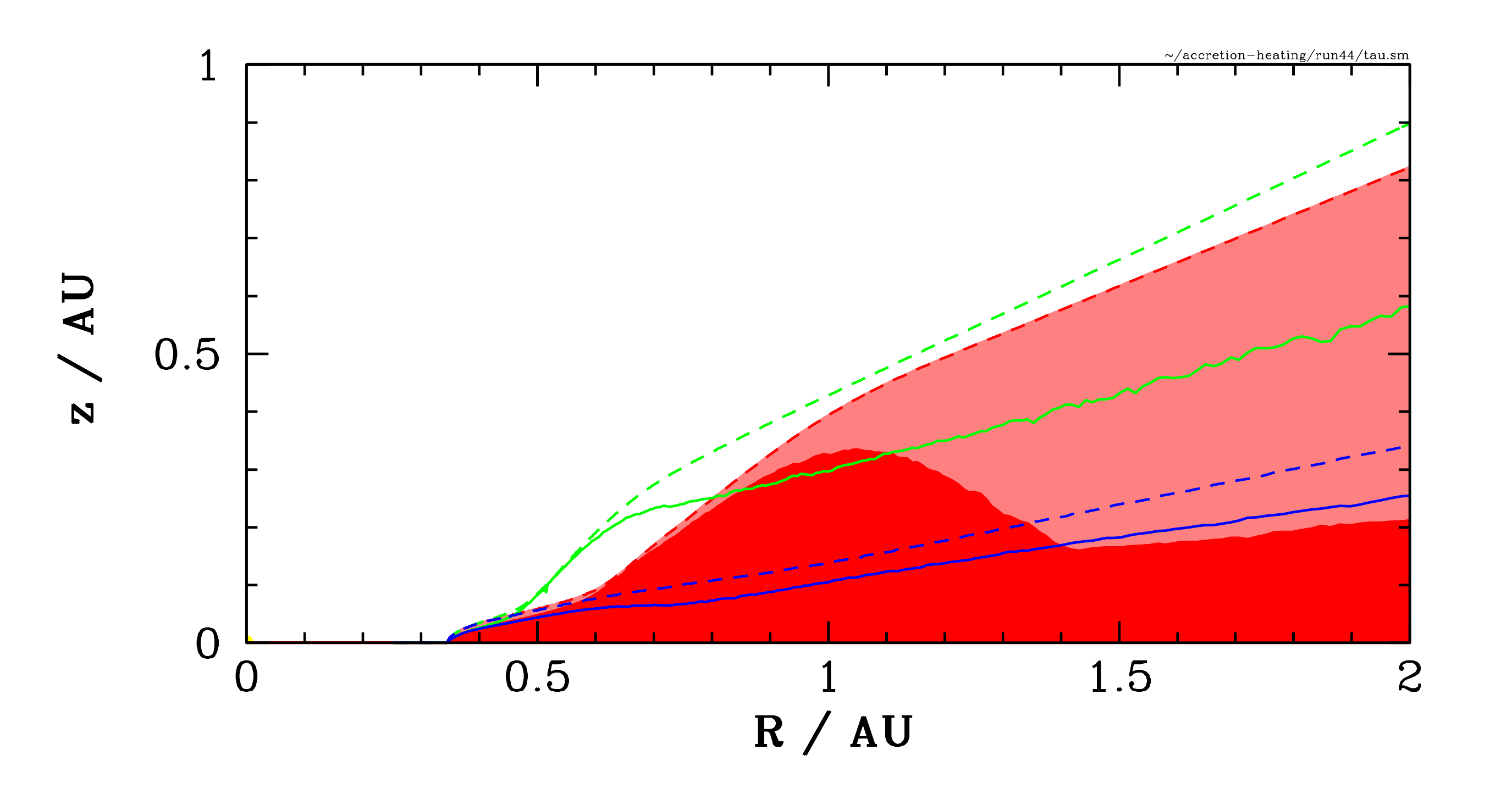}\\
   \includegraphics[width=0.58\linewidth]{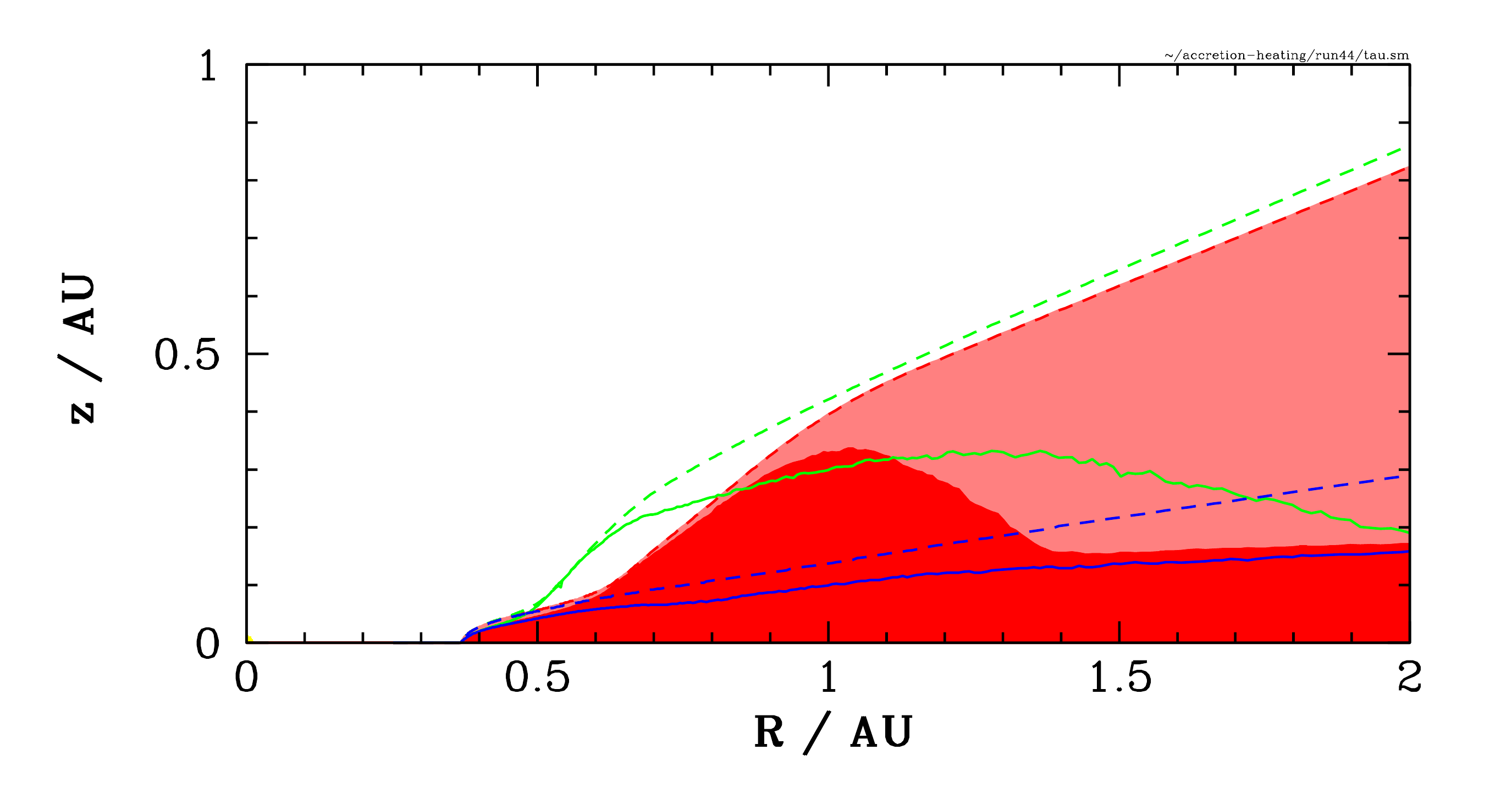}

   \figcaption{\sf Surfaces of unit optical depth for starlight
     entering the nine model disks.  The wavelength is 0.314~$\mu$m,
     near the starlight peak.  The three dust-depleted disks are at
     top, the three dusty disks at center and the three disks with
     dusty rings below.  In each panel the disk with gas support only
     is shown by blue lines, the disk with magnetic support throughout
     by green lines and the disk with the magnetically-supported bump
     by red lines and shading.  In each case the upper dashed curve is
     the surface of unit optical depth for photons arriving from the
     star, while the lower solid curve is the surface of unit optical
     depth for photons traveling vertically downward.  The yellow dot
     at the origin is the star to scale.
    \label{fig:tau3}}
\end{figure}

\begin{figure}[tb!]
  \centering
  \includegraphics[width=4cm]{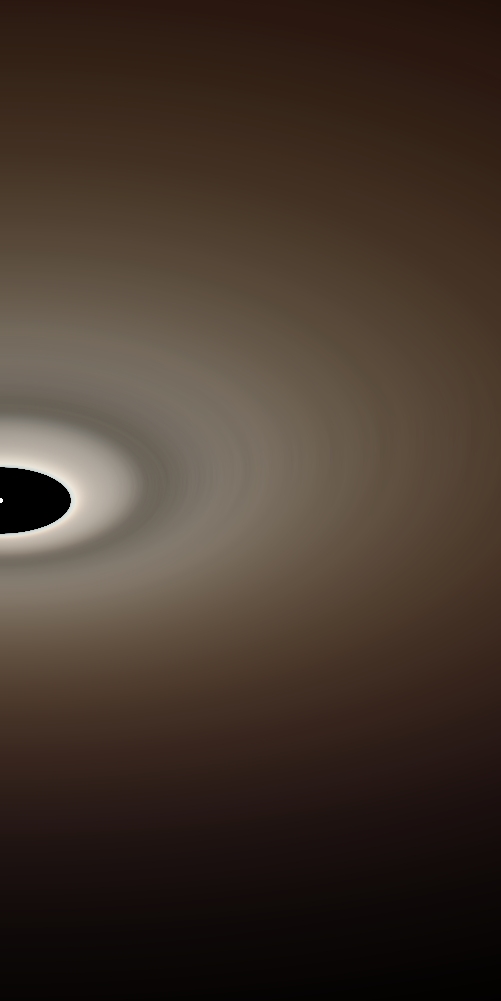}
  \includegraphics[width=4cm]{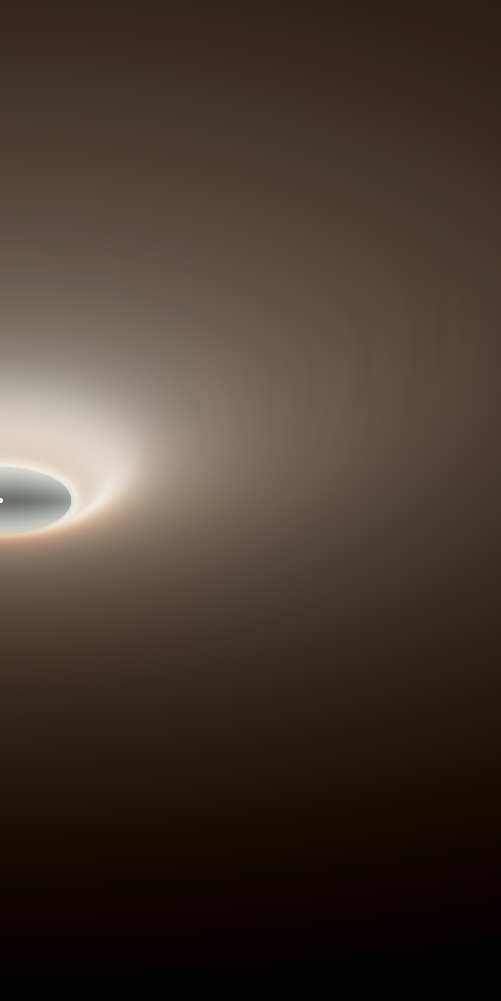}
  \includegraphics[width=4cm]{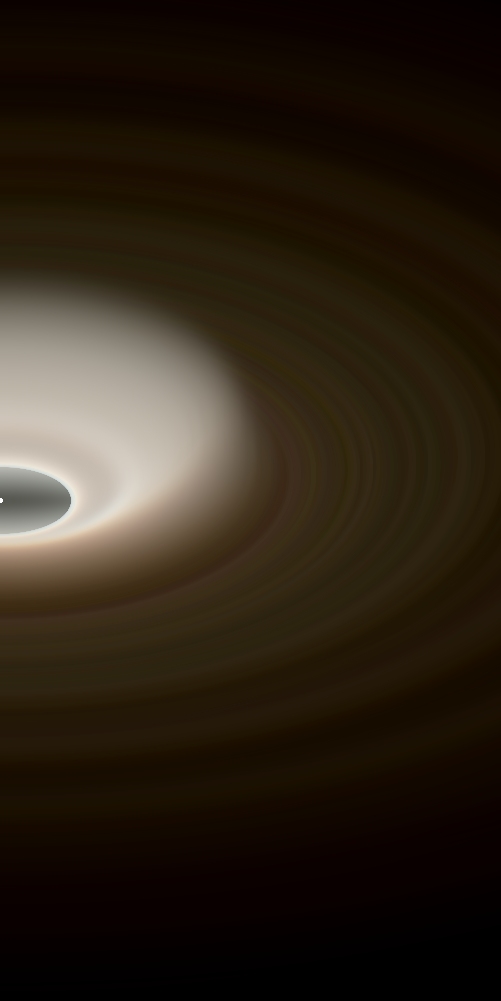}
  \figcaption{\sf Synthetic images of the central region of the dusty
    Herbig disk with (left to right) no magnetic support, magnetic
    support throughout, and a magnetically-supported bump.  The field
    of view is 2.5~AU wide and the system is inclined $60^\circ$ from
    face-on.  The star is shown to scale at the center of each panel's
    left edge.  The blue, green and red channels in each image
    correspond to wavelengths 1.25, 1.6 and 2.2~$\mu$m or J, H and
    K~bands, respectively.  A shared logarithmic intensity scale is
    used in all three panels.
    \label{fig:images}}
\end{figure}

The magnetically-supported atmosphere contains only a small fraction
of the disk mass.  A natural question to start with is therefore
whether the atmosphere is optically-thick enough to affect the
reprocessing of the starlight.  In figure~\ref{fig:tau3} we draw the
surfaces of unit starlight optical depth in the nine models.  The
dust-depleted cases in the first panel show clear differences between
versions with and without the magnetic support.  The
starlight-absorbing surface lies 1.75~times
higher at 1~AU with magnetic support throughout (green dashed line)
than in the hydrostatic version (blue dashed line).  The ratio is
1.48~times for the magnetically-supported bump (red dashed line).  The
magnetic support's effects are even stronger in the dusty cases
(second panel) where the corresponding ratios are~3.07 and~2.82.  The
magnetic support makes the starlight-absorbing surface taller by
similar factors in the cases with dusty inner rings (third panel)
where the ratios to the hydrostatic version are~3.06 and 2.87.
Furthermore, synthetic images of the central 2.5~AU show the
magnetically-supported material noticeably alters the appearance in
the near-infrared J, H and K bands (figure~\ref{fig:images}),
increasing the surface area of bright material lying within 1~AU.
Considering all six scenarios with magnetic support, clearly the
atmosphere reprocesses significant stellar luminosity beyond that
intercepted by the hydrostatic models.

Closer examination shows a striking correspondence between the surface
brightness in fig.~\ref{fig:images} and the slope of the
starlight-absorbing surface in fig.~\ref{fig:tau3}.  In particular,
the brightest of the three dusty models just inside 1~AU is the
magnetically-supported bump, which has its surface tilted most steeply
toward the star.

\begin{figure}[tb!]
   \centering
   \includegraphics[width=0.455\linewidth]{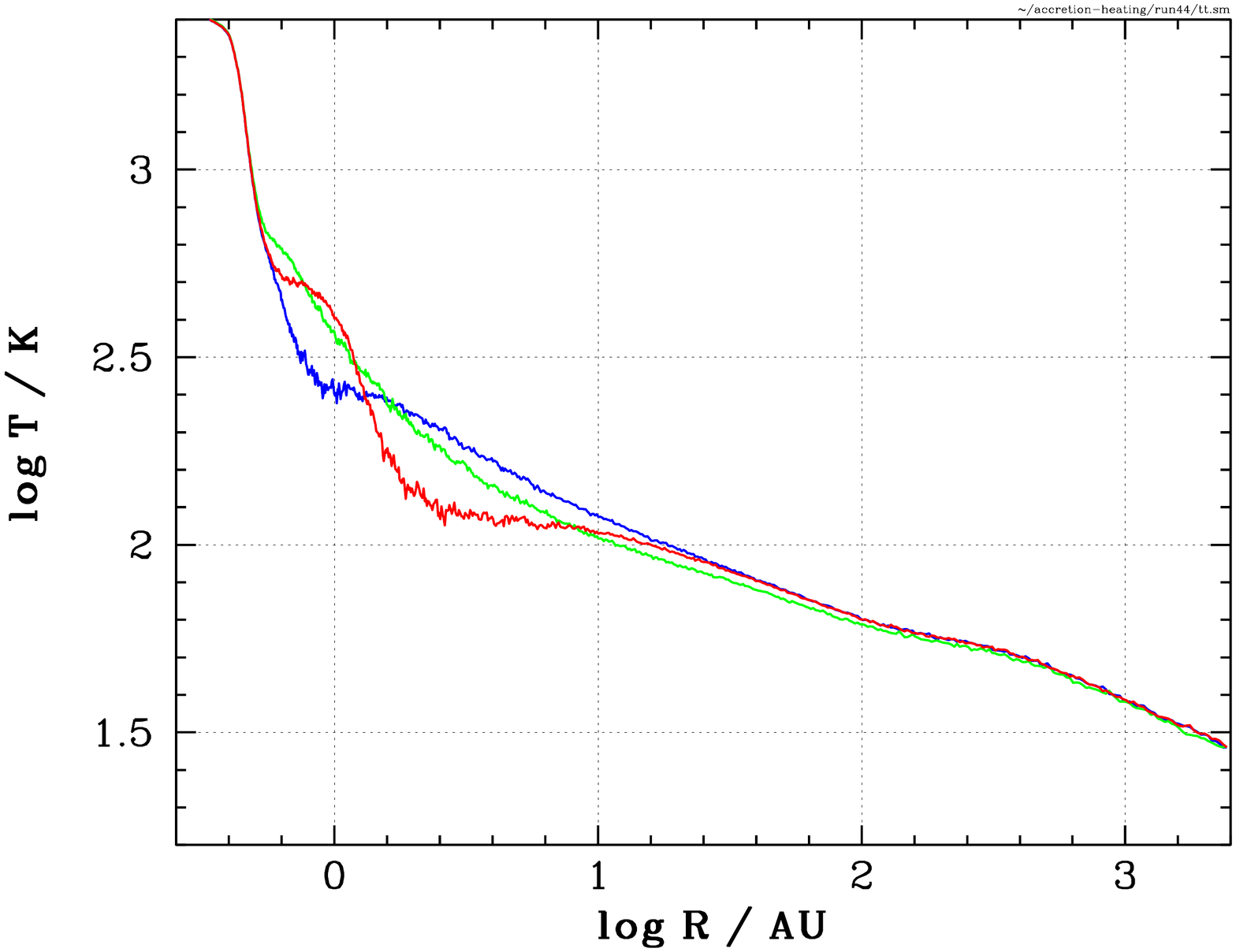}\\
   \includegraphics[width=0.455\linewidth]{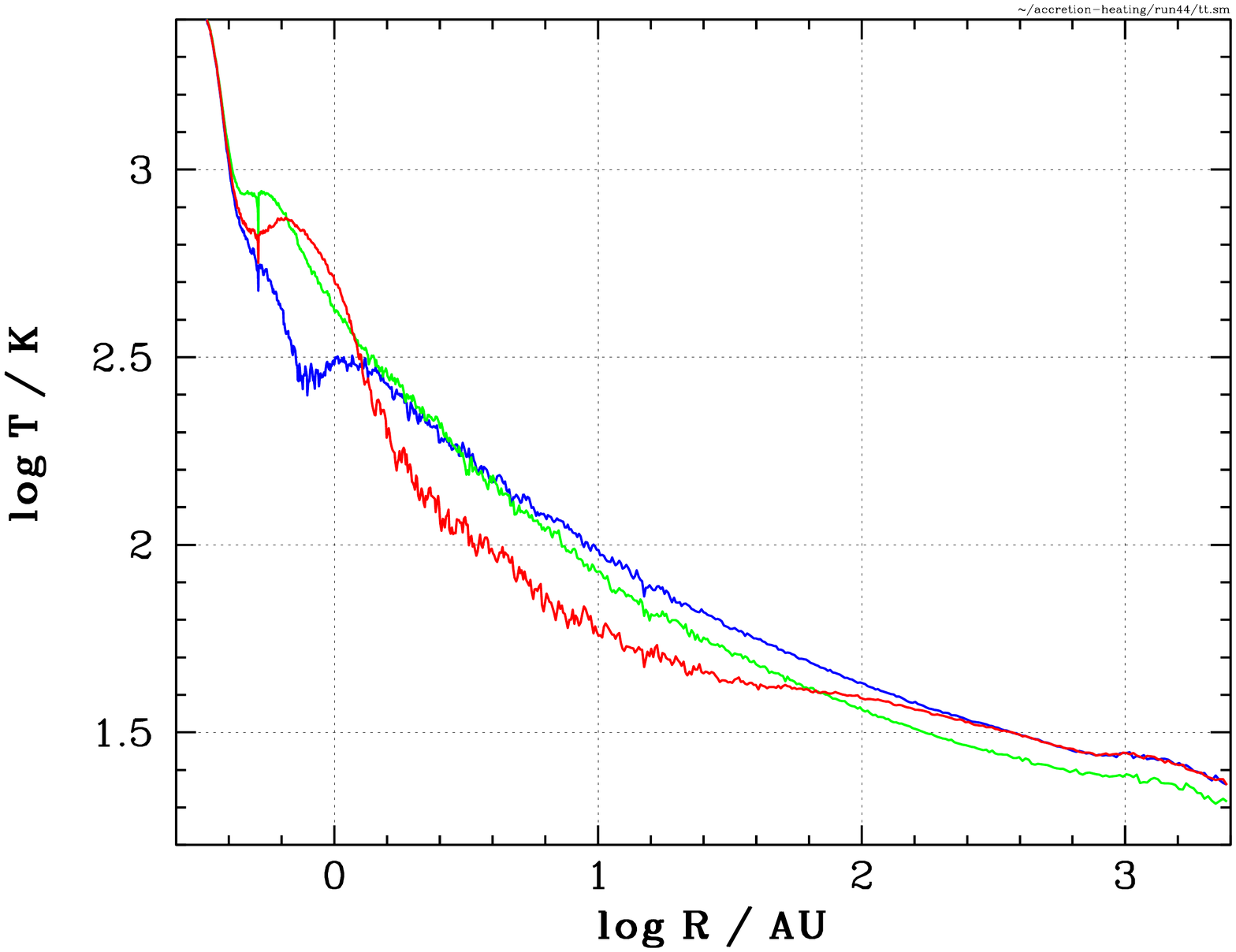}\\
   \includegraphics[width=0.455\linewidth]{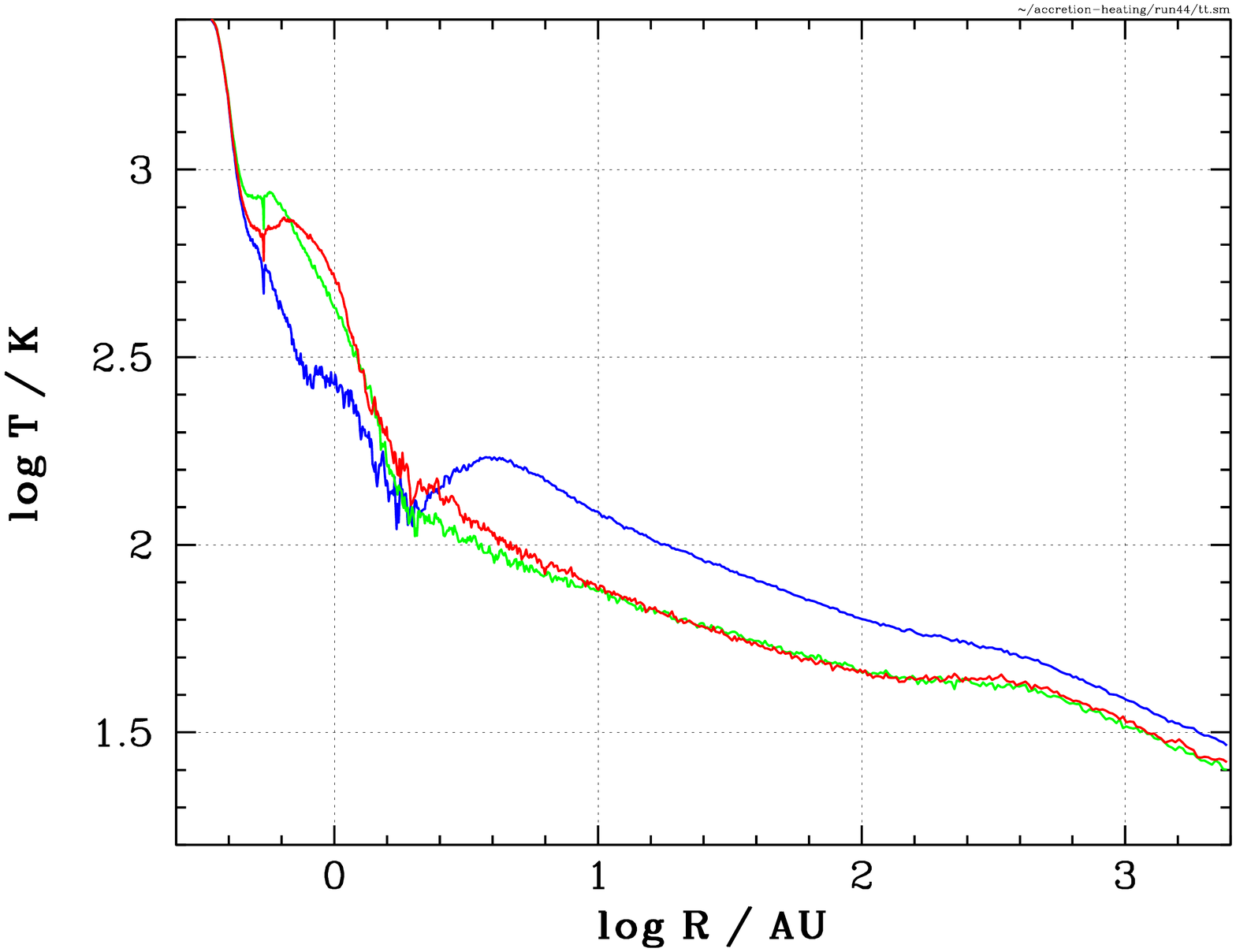}

   \figcaption{\sf Midplane temperature vs.\ radius in the nine model
     disks.  The dust-depleted disks are in the top panel, the dusty
     disks at center and the disks with dusty rings below.  Colors are
     as in figure~\ref{fig:tau3}: blue shows the disk with gas support
     only, green the disk with magnetic support throughout and red the
     disk with the magnetically-supported bump.
    \label{fig:ttvsr3}}
\end{figure}

Part of the extra height in the models with magnetically-supported
atmospheres comes directly from the magnetic support.  Another part
comes from a secondary effect: since the atmosphere intercepts more
starlight, the interior is hotter than in the hydrostatic models, and
the gas pressure scale height is greater.  The higher midplane
temperatures near 1~AU resulting from magnetic support can be seen in
figure~\ref{fig:ttvsr3}.  On the other hand, the
magnetically-supported bump casts a shadow that appears as a dip in
the temperature profiles at distances of several~AU.  Even the
hydrostatic model has a shadow when the dust abundance is greatest in
a ring near the disk's inner edge (figure~\ref{fig:ttvsr3} bottom
panel, blue curve).  Starlight grazing the top of the dusty ring
subsequently passes through less-opaque material, reaching unit
optical depth only near 4~AU.

\subsection{Is the System Bright Enough at Near-Infrared
  Wavelengths?\label{sec:seds}}

\begin{figure}[tbh!]
   \centering
   \includegraphics[width=0.40\linewidth]{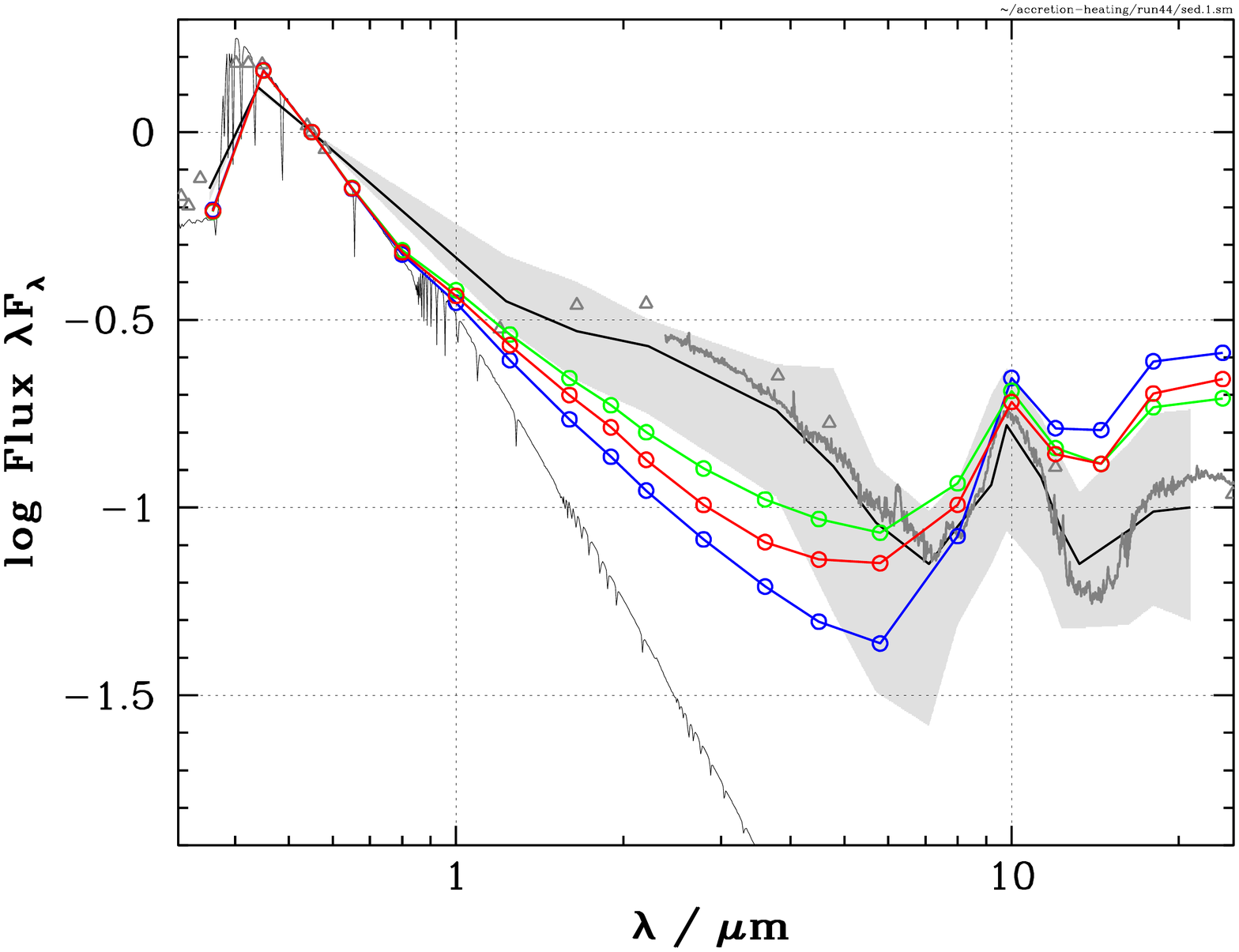}\\
   \includegraphics[width=0.40\linewidth]{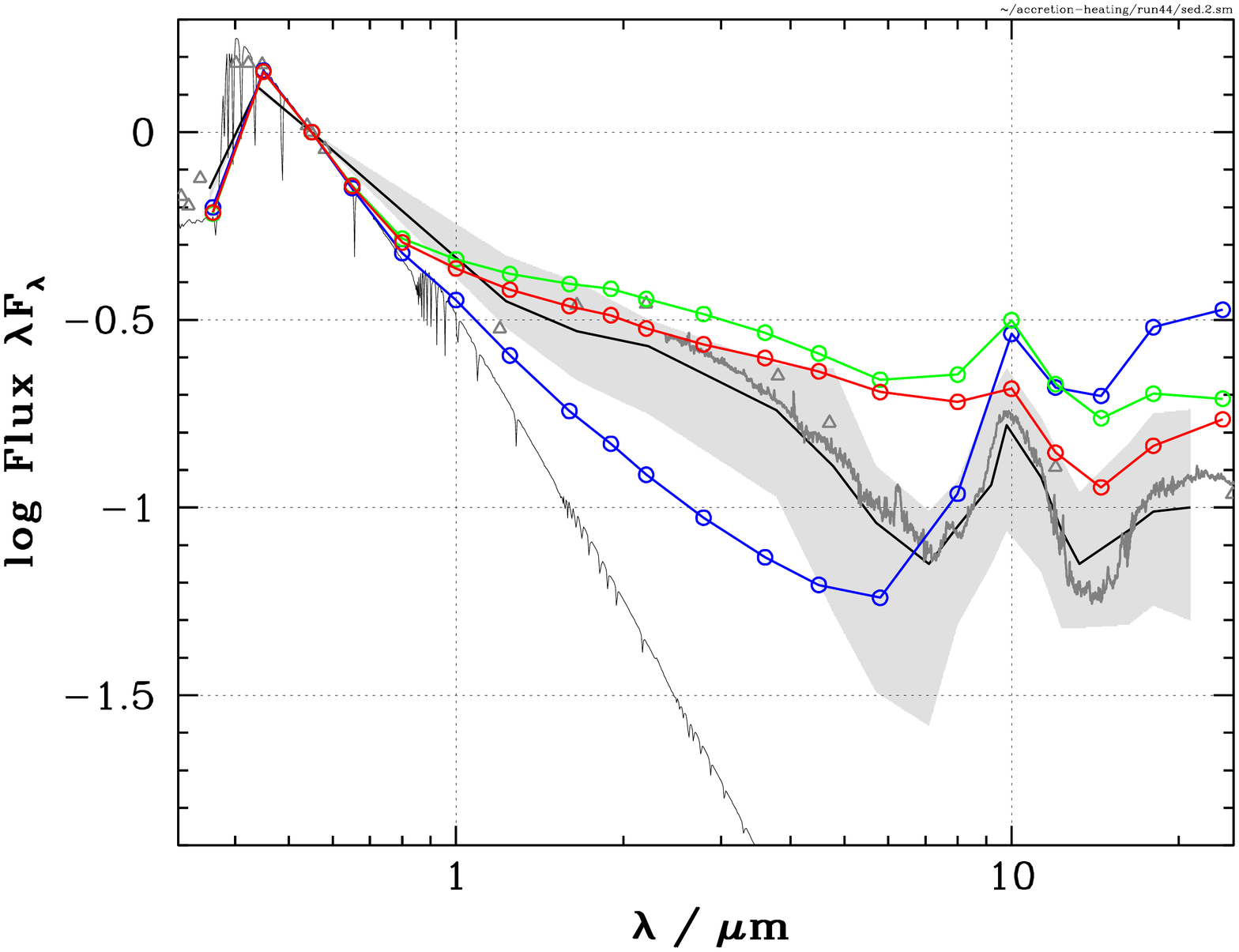}\\
   \includegraphics[width=0.40\linewidth]{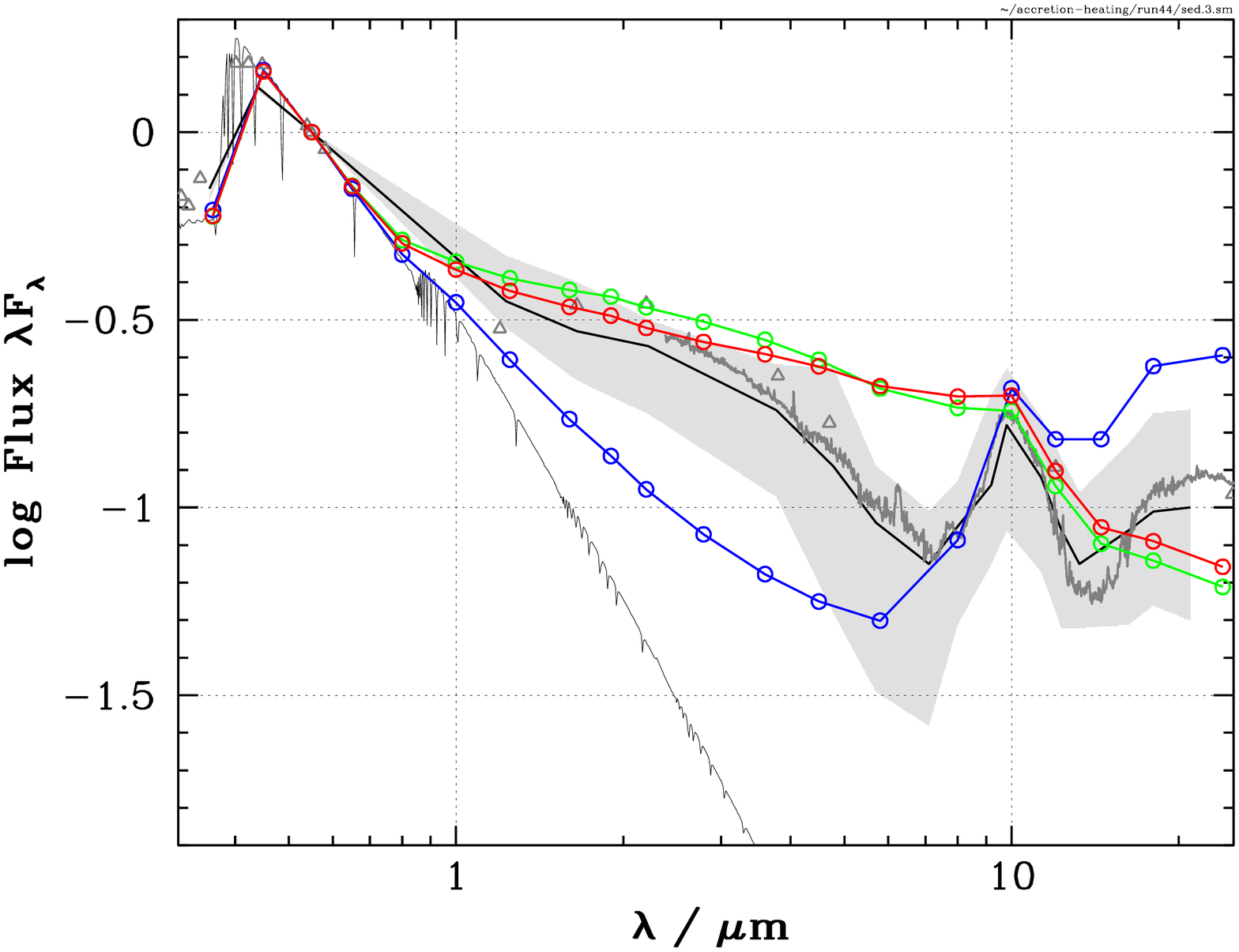}
   \figcaption{\sf Spectral energy distributions of the nine model
     Herbig systems.  The dust-depleted disks are at top, the dusty
     disks at center and the disks with dusty rings below.  In each
     panel the lines' colors are as in figure~\ref{fig:tau3}: blue
     is the disk with gas support only, green the disk with magnetic
     support throughout and red the disk with the
     magnetically-supported bump.  The thin black line indicates the
     stellar spectrum, a Kurucz model.  The thick black line marking
     the median Herbig SED falls in a light-gray band reaching from
     the first to the third quartile \citep{2012A&A...539A...9M}.  The
     dark grey curve and triangles indicate a spectrum and photometry
     of the star AB~Aurigae.  All are normalized at 0.55~$\mu$m.
    \label{fig:sed3}}
\end{figure}

We wish to know whether the magnetic support increases the
near-infrared excess enough to account for the observed SEDs.  From
the three spectral energy distributions in the first panel of
figure~\ref{fig:sed3}, we see that all the dust-depleted models are
too faint at wavelengths 2--4~$\mu$m by factors of two or more
relative to the median Herbig system.  Similar problems afflict the
hydrostatic models that are dusty throughout, as well as those that
are dusty only in a central ring (blue curves in second and third
panels).  By contrast, the versions with magnetic support lie close to
the median SED at near-infrared wavelengths (green and red curves).
Considering all nine models together, we see that four exceed the
median observed near-infrared excess.  All four have, just outside the
sublimation radius, both a magnetized atmosphere and a
near-interstellar dust-to-gas mass ratio.  The key to reprocessing the
extra starlight is a sufficient column of dust in the atmosphere, so
lower dust-to-gas ratios would allow a similar outcome if the disk
contained more gas than our chosen model (figure~\ref{fig:sd}).  With
enough dust present, the magnetic support readily accounts for the
excess that is missing from hydrostatic models.

The magnetically-supported models all do a poor job of matching the
10-$\mu$m silicate feature's steep short-wavelength side.  The most
likely reason is an incorrect shape for the starlight-absorbing
surface, due to the simple choices we made for the profiles of surface
density and magnetic support.  The disk annuli near 0.5~AU do show
synthetic spectra with suitably steep slopes from 6--10~$\mu$m,
suggesting that the median silicate band shape might be better matched
with a shorter magnetically-supported bump placed nearer the star.  A
further possibility is that typical Herbig disks' opacities have a
wavelength dependence differing from the curves we used.  Testing
these ideas is a challenge for the future.  Each calculation takes
about a week of computer time and many parameters remain to be varied.

At wavelengths beyond 15~$\mu$m, the flux and SED slope are affected
by the shape of the disk surface at and outside 10~AU.  Magnetic
support generally means more starlight intercepted near 1~AU and less
outside 10~AU, making the outer annuli cooler and leading to steeper
declines in flux with wavelength.  The anti-correlation between the
7-$\mu$m excess and the 13.5-to-7-$\mu$m flux ratio observed by
\citet{2009A&A...502L..17A} can thus qualitatively be explained by a
variation from one system to the next in the strength of the magnetic
support.

On the near-infrared bump's other side, from 2~$\mu$m shortward, there
is interferometric evidence for emission arising within the
sublimation radius in some systems \citep{2007ApJ...657..347E,
  2008A&A...483L..13I, 2010A&A...511A..74B, 2010ApJ...718..774E}.  Our
radiative transfer modeling does not address this component of the
system.  However if, as we propose, the near-infrared bump arises in
material supported by the same magnetic fields that drive accretion,
then correlations might be expected between the bump's height, the
surface density of the material within the sublimation radius, and the
accretion signatures such as H$\alpha$ emitted near the stellar
photosphere.  Simultaneous optical and infared observations could help
illuminate such a connection.

\subsection{Can Grains Remain Suspended in the Atmosphere?
  \label{sec:settling}}

We have assumed the grains providing the opacity are well-mixed in the
gas.  This is valid if the grains are stirred up, either by the
turbulence or by the magnetic buoyancy, faster than they settle.

Magneto-rotational turbulence can loft material no quicker than the
velocity correlation timescale, which is a fraction of the orbital
period \citep{2006A&A...452..751F}.  The linear magneto-rotational
instability is slow-growing or stabilized high in the atmosphere,
where the plasma beta is less than unity \citep{2000ApJ...540..372K}.
Even so, in non-linear stratified shearing-box MHD calculations the
velocity dispersion remains large at these heights
\citep{2000ApJ...534..398M, 2010MNRAS.409.1297F, 2011ApJ...742...65O}.
However out of an abundance of caution we set the stirring timescale
to the slower magnetic buoyancy timescale, measured using the
``butterfly'' pattern visible when the magnetic pressure is plotted
versus height and time.  In shearing-box calculations this timescale
is around ten orbits, with or without a dead zone
\citep{2000ApJ...534..398M, 2010MNRAS.409.1297F, 2011ApJ...732L..30H}.
The upshot is that the grains repopulate the atmosphere within ten
orbits if sufficiently coupled to the gas.

On the other hand, settling removes grains from the atmosphere with a
speed such that the drag force balances the vertical component of the
star's gravity.  The drag force is in the Epstein regime where the gas
molecules' mean free path exceeds the grain size.  The settling time
is the distance to the midplane divided by the settling speed and for
compact spherical particles is given by
\begin{equation}
  {t_{\rm sett}\over t_K} = {1\over 4\pi^2} {t_K\over t_{\rm drag}},
\end{equation}
where $t_K$ is the Keplerian orbital period, and the drag stopping
time $t_{\rm drag} = {\rho_d a/(\rho c_s)}$ depends on the grains'
internal density $\rho_d$ and radius $a$ using the notation from
\citet{2010ApJ...708..188T}.

Now the grains to be concerned with are those that absorb the
starlight and give off the disk inner rim's thermal infrared emission.
The starlight peaks near wavelength 0.3~$\mu$m, the infrared emission
near 3~$\mu$m.  The biggest contribution to the opacity is from grains
with circumference comparable to the wavelength
\citep{1957lssp.book.....V}.  Most important for the starlight opacity
are thus grains smaller than $a=0.1$~$\mu$m, and for the infrared
opacity grains around 0.5~$\mu$m in radius.  In discussing the
settling of these particles we take an internal density
$\rho_d=3$~g~cm$^{-3}$, similar to that of terrestrial basalt.
Densities are less for carbon-rich grains.

\begin{figure}[tb!]
  \centering
  \includegraphics[width=0.58\linewidth]{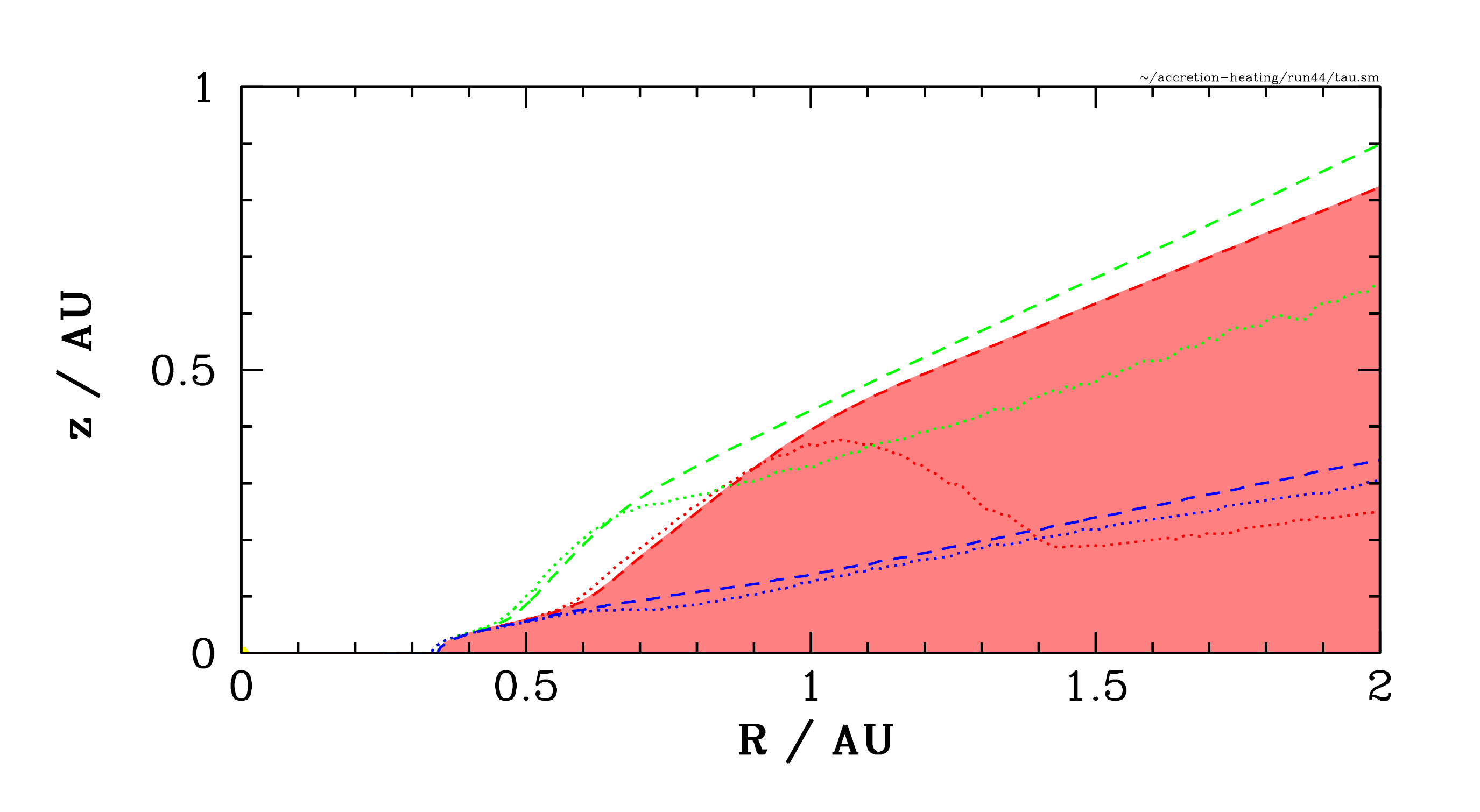}\\
  \includegraphics[width=0.58\linewidth]{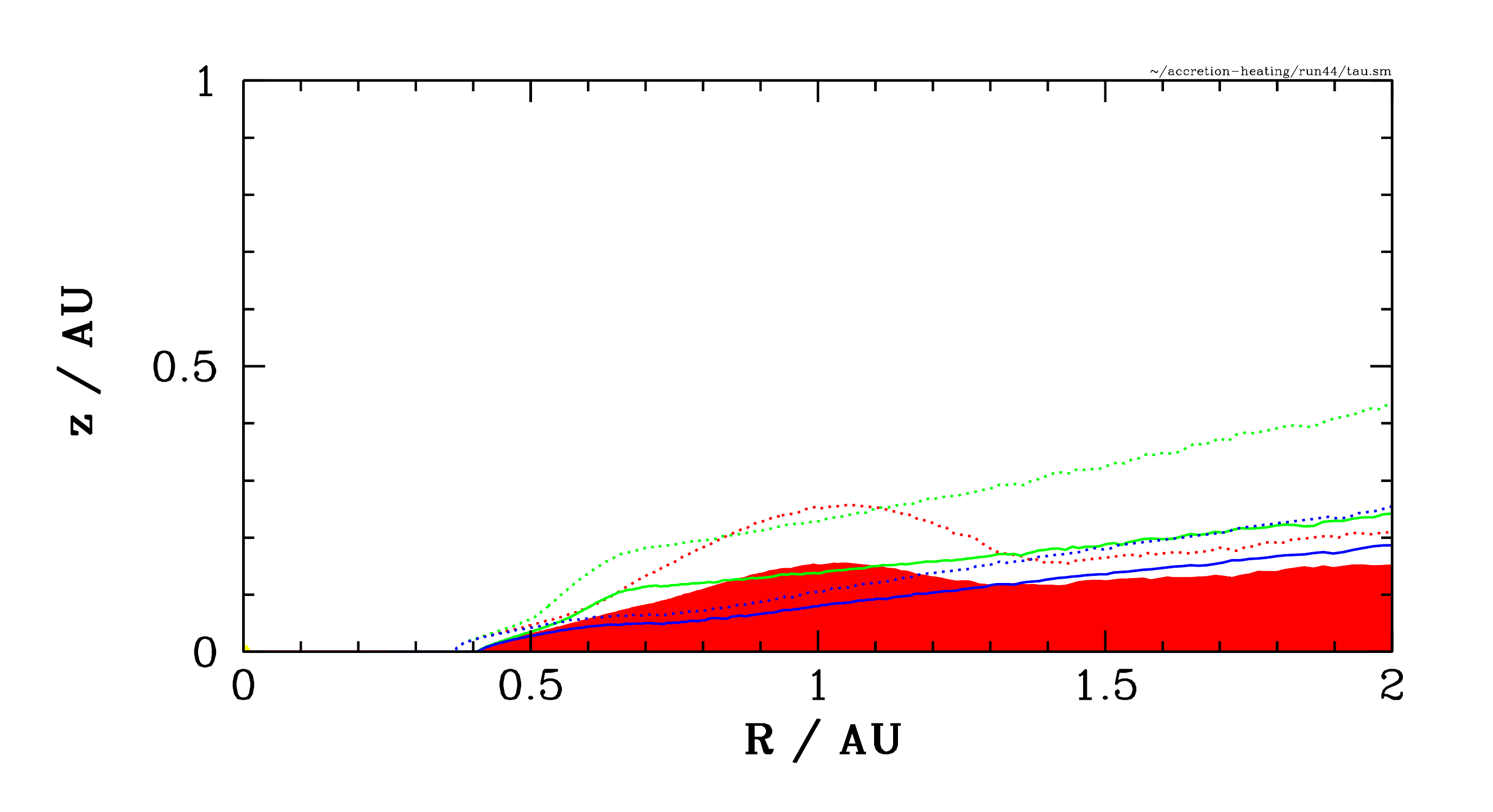}

  \figcaption{\sf {\it Upper panel:} Dotted lines mark the height
    above which the settling time is less than ten orbits for
    0.1-$\mu$m grains in the three dusty disks.  The dashed lines show
    the starlight-absorbing surface with the grains well-mixed, and
    are reproduced from figure~\ref{fig:tau3} middle panel.  The disk
    with gas support only is blue, with magnetic support throughout is
    green and with the magnetically-supported bump is the red line and
    shading.  In all three cases, settling can lower the
    starlight-absorbing surface only slightly.  {\it Lower panel:}
    Corresponding settling surfaces for 1-$\mu$m grains (dotted),
    together with the unit vertical optical depth surfaces for thermal
    infrared emission (solid lines and shading).  The same three dusty
    models appear in the same colors as in the upper panel.  The
    settling line lies well above the infrared-emitting surface
    throughout, indicating settling is unimportant at the infrared
    photosphere.
    \label{fig:settle}}
\end{figure}

On the top panel of figure~\ref{fig:settle} we show by dotted lines
the height to which the 0.1-$\mu$m grains settle within ten orbits.
The settling heights are overlaid on the surfaces of unit direct
starlight optical depth for the dusty models from
figure~\ref{fig:tau3}.  We see that settling is important only in the
uppermost reaches of the atmosphere.  For each of the three disk
configurations, the dotted line lies above the unit-optical-depth
curve on the part of the disk most directly facing the star.  If
settling were to remove all grains above the dotted line, an unlikely
prospect, the height of the starlight-absorbing surface at 2~AU would
be reduced about 11\% in the hydrostatic disk, 27\% in the disk with
magnetic support throughout and 10\% in the disk with the
magnetically-supported bump.  Note that the starlight-absorbing height
beyond the bump is determined by the optically-thick bump itself.
Furthermore, the settling would be unimportant right up to the
starlight-absorbing surface in all the models if the inner disk had
gas surface densities a few times greater than we have assumed.

The situation for the largest grains contributing to the opacity
curves is shown in figure~\ref{fig:settle} bottom panel.  Particles
1~$\mu$m in radius settle significantly only in gas lying well above
the 3-$\mu$m photosphere.

The dusty models shown in figure~\ref{fig:settle} impose the most
severe settling constraint because their high opacity means the
starlight is absorbed in gas with the lowest density.  In the
corresponding plots for the three dust-depleted disks, not shown, the
starlight-absorbing surface falls well below the height where settling
begins to matter.  We conclude that the grains contributing most to
the starlight opacity in our model disks are never more than
marginally affected by settling, while the grains important for the
thermal emission are unaffected.

The Lorentz force plays a minor role in the grains' movements even in
the disk with magnetic support throughout.  Consider a grain traveling
at its settling speed through a magnetic field with pressure ten times
the local gas pressure.  Struck by electrons from the surrounding
plasma, the grain charges to the Coulomb limit in which the electric
repulsion reduces the cross-section for colliding with further
electrons to the point where the grain receives electrons and
slower-moving ions at equal rates.  The mutual electric potential
between the grain and an electron approaching its surface is then
about three times the thermal energy, as we determine by solving
\citet{2009ApJ...698.1122O} eq.~35.  Grains 0.1~$\mu$m in radius near
the silicate sublimation front charge to about 20~electrons.  Under
these conditions we find that the Lorentz force exceeds the
gravitational and drag forces only for grains located at and above the
uppermost, dashed green line in figure~\ref{fig:settle}.

\section{SUMMARY AND CONCLUSIONS\label{sec:conclusions}}

Many young intermediate-mass stars show near-infrared excesses that
have proven too large to explain using hydrostatic models.  The
hydrostatic disks are geometrically-thin near the silicate sublimation
radius, intercepting and reprocessing too small a fraction of the
starlight.  On the other hand, MHD calculations indicate that
magnetically-supported atmospheres are a generic feature of
protostellar disk annuli undergoing accretion driven by
magneto-rotational turbulence.  The MHD results show that magnetic
forces suspend small amounts of material well above the hydrostatic
photosphere.  To see whether such an atmosphere can account for the
excess near-infrared emission, we added simple exponential atmospheres
to global models of the disk around a Herbig star.  We placed the
models jointly in vertical magnetohydrostatic balance and in radiative
equilibrium with the starlight using Monte Carlo transfer
calculations, and constructed synthetic observations of the resulting
structures.  Our main findings are that (1) the atmosphere near the
sublimation radius is optically-thick to the starlight, and (2) if the
dust abundance there is close to the interstellar value, the resulting
near-infrared excess is sufficient to explain the median Herbig SED.
We therefore suggest that magnetically-supported atmospheres are a
common feature of at least the thermally-ionized inner annuli in
Herbig disks.

Magnetically-supported disk atmospheres have several further
interesting implications.  First, the disk thickness measures not the
temperature as long assumed, but the magnetic field strength.  The gas
pressure support yields only a lower bound on the height of the
starlight-absorbing surface.  The disk thickness is thus a gauge of
the strength of the magnetic activity.  Second, the disk atmosphere is
quite diffuse.  The exponential density profile falls off slower with
height than the traditional Gaussian.  The extended low-density
atmosphere offers a natural explanation for the finding from
interferometry, that the visibilities are better fit by hydrostatic
disks combined with spherical or near-spherical halos than by either
component alone.  Kinematic diagnostics will be valuable in the future
to help distinguish whether the extended material forms a turbulent
magnetized atmosphere, an escaping wind, or perhaps most likely, both.
Third, the inner disk's atmosphere can throw a shadow across the
material outside.  A flared shape may bring the distant parts of the
disk back up into the starlight.  Dips in the radial surface
brightness profile then do not have to be surface density deficits,
but could be the shadows cast at sunset by magnetically-supported
hills.  Fourth, the atmosphere can obscure our view of the star even
when the system is viewed at moderate inclination.  Accounting for
time-variable circumstellar extinction will be easier with magnetic
fields holding some gas and dust aloft.  In considering the
consequences for the extinction, it is worth noting that we took a
horizontally- and time-averaged atmospheric density profile, ignoring
the structure that is universally part of magnetic activity in MRI
turbulence as well as other contexts such as the Solar chromosphere.

\acknowledgements

We gratefully acknowledge discussions with C.\ Dominik, M.\ Flock,
G.\ Mulders and A.\ Natta.  The research was carried out in part at
the Jet Propulsion Laboratory, California Institute of Technology,
under a contract with the National Aeronautics and Space
Administration and with the support of the NASA Origins of Solar
Systems program via grant 11-OSS11-0074.  NJT was also supported by
the Alexander von Humboldt Foundation through a Fellowship for
Experienced Researchers.  SH was supported by JSPS KAKENHI grants
numbers 24540244 and 23340040.  Copyright 2013 California Institute of
Technology.  Government sponsorship acknowledged.

\bibliographystyle{apj}
\bibliography{ysodisk}

\end{document}